\documentclass{eas}
\usepackage{graphicx}

\begin{document}


\title{Blazhko Effect in Double Mode Cepheids}

\author{P. Moskalik}\address{Copernicus Astronomical Center, Bartycka 18, 00-716 Warsaw, Poland}
\author{Z. Ko{\l}aczkowski}\address{Universidad de Concepcion, Departamento de Fisica,
                                    Casilla 160-C, Concepcion, Chile}

\begin{abstract}
Systematic survey for multiperiodicity in the LMC Cepheids
(Moskalik, Ko{\l}aczkowski \& Mizerski \cite{MKM04}, \cite{MKM06})
has led to discovery of several new forms of pulsational
behaviour. One of them is periodic amplitude and phase modulation
observed in many first/second overtone (FO/SO) double mode
Cepheids. In the current paper we present detailed discussion of
this newly discovered phenomenon, based on a combined OGLE+MACHO
sample of double mode pulsators.
\end{abstract}

\maketitle


\section{Time Series Analysis}

We searched for additional signal in the data using a standard
consecutive pre\-white\-ning technique. To that effect, we first
fitted the data with the double frequency Fourier sum representing
pulsations in two radial modes:

\vskip 0pt

$$m(t) = \langle m\rangle + \sum_{j,k} {\rm A}_{jk} \sin [2\pi(j{\rm f}_1 + k{\rm f}_2)t + \phi_{jk}].\eqno(1)$$

\smallskip

\noindent The frequencies of the modes, ${\rm f}_1$ and ${\rm f}_2$,
were also optimized. The residuals of the fit were then
searched for additional periodicities. This was done with the
Fourier transform, calculated over the range of $0-5$\thinspace c/d. In
the next step, a new Fourier fit with all frequencies discovered
so far was performed and the fit residuals were searched for
additional frequencies again. The process was stopped when no new
frequencies were detected.

\section{LMC Sample}

\begin{table}
\caption{Blazhko FO/SO Double-Mode Cepheids in the LMC}
\begin{center}
\footnotesize
\begin{tabular}{lcccccccccc}
\hline
\noalign{\smallskip}
             &        &        &         &       & $A_1^{-}$
                                                         & $A_1^{- -}$
                                                                 &       & $A_2^{-}$
                                                                                 & $A_2^{- -}$
                                                                                         \\
\noalign{\smallskip}
OGLE ID/     & ${\rm P}_1$
                      & ${\rm P}_2$
                               & ${\rm P}_{\rm B}$
                                         & $A_1$ & $A_1^{+}$
                                                         & $A_1^{++}$
                                                                 & $A_2$ & $A_2^{+}$
                                                                                 & $A_2^{++}$
                                                                                         \\
\noalign{\smallskip}
~/MACHO ID   & [day]  & [day]  &  [day]  & [mag] & [mag] & [mag] & [mag] & [mag] & [mag] \\
\noalign{\smallskip}
\hline
\noalign{\smallskip}

SC1--44845   & 0.9520 & 0.7660 & ~~794.0 & 0.165 & 0.021 &  ---  & 0.046 & 0.021 &  ---  \\
             &        &        &         &       & 0.011 &  ---  &       & 0.010 &  ---  \\

\noalign{\smallskip}

SC1--285275  & 0.8566 & 0.6892 & ~~891.6 & 0.177 & 0.031 &  ---  & 0.040 & 0.022 & 0.008 \\
             &        &        &         &       & 0.014 &  ---  &       & 0.013 &  ---  \\

\noalign{\smallskip}

SC1--335559  & 0.7498 & 0.6036 & ~~779.2 & 0.227 & 0.019 &  ---  & 0.062 & 0.023 &  ---  \\
             &        &        &         &       &  ---  &  ---  &       &  ---  &  ---  \\

\noalign{\smallskip}

SC2--55596   & 0.9325 & 0.7514 & ~~768.2 & 0.146 & 0.007 &  ---  & 0.039 &  ---  &  ---  \\
             &        &        &         &       & 0.007 &  ---  &       &  ---  &  ---  \\

\noalign{\smallskip}

SC6--142093  & 0.8963 & 0.7221 &  1101.6 & 0.154 & 0.028 &  ---  & 0.043 & 0.017 & 0.012 \\
             &        &        &         &       & 0.013 &  ---  &       & 0.017 & 0.006 \\

\noalign{\smallskip}

SC6--267410  & 0.8885 & 0.7168 & ~~856.9 & 0.120 &  ---  &  ---  & 0.036 &  ---  &  ---  \\
             &        &        &         &       &  ---  &  ---  &       & 0.033 &  ---  \\

\noalign{\smallskip}

SC8--10158   & 0.6900 & 0.5557 &  1060.7 & 0.175 & 0.011 &  ---  & 0.031 & 0.019 &  ---  \\
             &        &        &         &       &  ---  &  ---  &       & 0.012 &  ---  \\

\noalign{\smallskip}

SC11--233290 & 1.2175 & 0.9784 &  1006.2 & 0.186 & 0.019 &  ---  & 0.043 & 0.017 & 0.006 \\
             &        &        &         &       & 0.010 &  ---  &       & 0.010 &  ---  \\

\noalign{\smallskip}

SC15--16385  & 0.9904 & 0.7957 &  1123.1 & 0.258 & 0.017 &  ---  & 0.049 & 0.020 &  ---  \\
             &        &        &         &       & 0.016 &  ---  &       &  ---  &  ---  \\

\noalign{\smallskip}

SC20--112788 & 0.7377 & 0.5943 &  1379.2 & 0.164 & 0.062 & 0.022 & 0.015 & 0.016 &  ---  \\
             &        &        &         &       & 0.019 &  ---  &       & 0.013 &  ---  \\

\noalign{\smallskip}

SC20--138333 & 0.8598 & 0.6922 & ~~795.0 & 0.189 & 0.013 &  ---  & 0.060 & 0.018 &  ---  \\
             &        &        &         &       & 0.008 &  ---  &       & 0.010 &  ---  \\

\noalign{\smallskip}

2.4909.67    & 1.0841 & 0.8700 &  1019.7 & 0.216 & 0.012 &  ---  & 0.055 & 0.013 &  ---  \\
             &        &        &         &       & 0.010 &  ---  &       & 0.009 &  ---  \\

\noalign{\smallskip}

13.5835.55   & 0.8987 & 0.7228 &  1074.9 & 0.244 & 0.037 & 0.010 & 0.040 & 0.026 & 0.012 \\
             &        &        &         &       & 0.021 &  ---  &       & 0.016 &  ---  \\

\noalign{\smallskip}

14.9585.48   & 0.9358 & 0.7528 &  1092.5 & 0.139 & 0.024 &  ---  & 0.036 & 0.025 & 0.012 \\
             &        &        &         &       & 0.008 &  ---  &       & 0.010 &  ---  \\

\noalign{\smallskip}

17.2463.49   & 0.7629 & 0.6140 &  1069.9 & 0.235 & 0.018 &  ---  & 0.051 & 0.013 &  ---  \\
             &        &        &         &       & 0.014 &  ---  &       & 0.015 &  ---  \\

\noalign{\smallskip}

18.2239.43   & 1.3642 & 1.0933 & ~~706.8 & 0.230 & 0.035 & 0.010 & 0.063 & 0.017 &  ---  \\
             &        &        &         &       & 0.014 &  ---  &       & 0.015 &  ---  \\

\noalign{\smallskip}

22.5230.61   & 0.6331 & 0.5101 & ~~804.3 & 0.221 &  ---  &  ---  & 0.045 & 0.016 &  ---  \\
             &        &        &         &       &  ---  &  ---  &       & 0.014 &  ---  \\

\noalign{\smallskip}

23.2934.45   & 0.7344 & 0.5918 & ~~797.6 & 0.207 & 0.023 &  ---  & 0.055 & 0.031 & 0.015 \\
             &        &        &         &       &  ---  &  ---  &       & 0.026 & 0.011 \\

\noalign{\smallskip}

23.3184.74   & 0.8412 & 0.6778 &  1126.0 & 0.160 & 0.015 &  ---  & 0.048 & 0.011 &  ---  \\
             &        &        &         &       &  ---  &  ---  &       & 0.013 &  ---  \\

\noalign{\smallskip}

80.7080.2618 & 0.7159 & 0.5780 & ~~920.3 & 0.168 & 0.010 &  ---  & 0.055 & 0.006 &  ---  \\
             &        &        &         &       &  ---  &  ---  &       & 0.009 &  ---  \\

\noalign{\smallskip}
\hline
\end{tabular}
\end{center}
\label{table1}
\end{table}

\begin{figure}[h]
\begin{center}
\includegraphics[width=11cm]{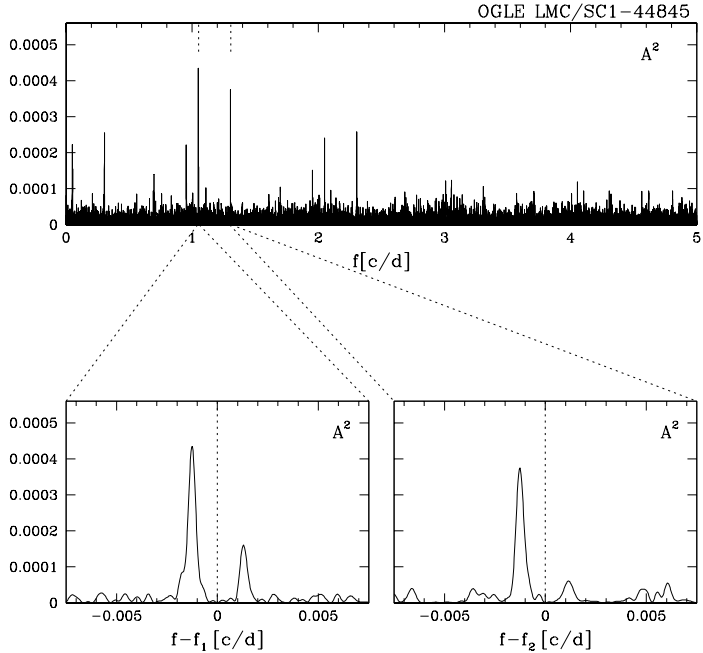}
\end{center}
\vskip -2.5truecm
\caption{Power spectrum of FO/SO double mode Cepheid
LMC/SC1--44845 after prewhitening with frequencies of the two
radial modes (and their linear combinations). Removed radial
frequencies are indicated by dashed lines. Lower panels display
the fine structure around the radial modes.}
\label{fig1}
\end{figure}

Our initial analysis was performed with DIA-reduced OGLE-II
photometry (\.Ze\-bru\'n {\em et~al.\/} \cite{ZEB01}). There are
56 FO/SO double mode Cepheids in the OGLE catalog (Soszy\'nski
{\em et~al.\/} \cite{SOSZ00}). In one-third of these stars we
detected residual signal close to one or both primary (radial)
pulsation frequencies. In all cases, however, the secondary
frequencies were not resolved from the primary ones. Clearly,
$\sim\! 1200$\thinspace days timebase provided by OGLE-II data was
insufficient for our purpose. Therefore, the final frequency
analysis was performed with MACHO photometry (Allsman \& Axelrod
\cite{AA01}), which had lower precision, but offered much longer
timebase ($2700-2800$\thinspace days).

The OGLE-II sample was supplemented by 51 additional FO/SO double
mode Cepheids discovered by MACHO team outside the area covered by
OGLE-II (Alcock {\em et~al.\/} \cite{MACHO99}, \cite{MACHO03}).
Our final FO/SO Cepheid sample consisted of 107 objects.

\section{Results}

Residual power close to the primary pulsation frequencies was
detected in 37 FO/SO double mode Cepheids (35\% of the sample). In
20 of these stars, which constitute 19\% of the entire sample, we
were able to resolve this power into indi\-vi\-dual frequencies. We
consider two frequencies to be resolved if $1/\Delta{\rm f} <
1400$\thinspace days. All resolved stars are listed in
Table\thinspace\ref{table1}.

\subsection{Frequency Domain}

\begin{figure}[t]
\begin{center}
\includegraphics[width=11cm]{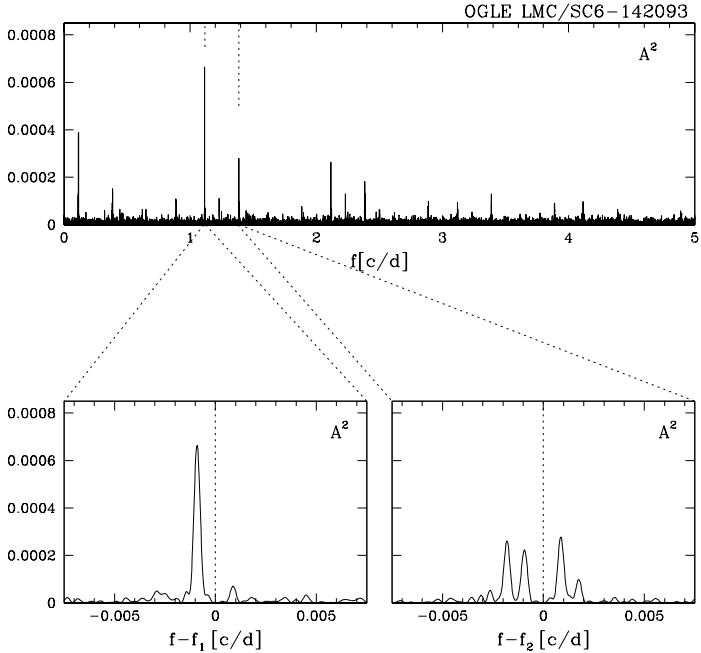}
\end{center}
\vskip -2.45truecm
\caption{Same as Fig.\thinspace\ref{fig1}, but for LMC/SC6--142093.}
\label{fig2}
\end{figure}

Resolved FO/SO double mode Cepheids display very characteristic
frequency pattern. In most cases, we detect two peaks in the
vicinity of each radial frequency. They are located on the
opposite sides of the primary (radial) peak and together with the
primary frequency they form an {\it equally spaced frequency
triplet}. In majority of cases, such a structure is detected
around {\it both} radial modes. Typical example of this pattern is
displayed in Fig.\thinspace\ref{fig1}. In two Cepheids we detect
{\it equally spaced frequency quintuplets} centered on the second
overtone. One of these objects is presented in
Fig.\thinspace\ref{fig2}. Incomplete quintuplets are also seen in
6 other stars.

The secondary peaks are always very small, with amplitudes of only
15\thinspace mmag on average and never higher than 62\thinspace
mmag. The separation of components in triplets/quintuplets,
$\Delta{\rm f}$, is also very small and never exceed
0.00142\thinspace c/d. This corresponds to the beat period of
${\rm P}_{\rm B} > 700$\thinspace day. The lower limit for
$\Delta{\rm f}$ (upper limit for ${\rm P}_{\rm B}$) is currently
given by resolution of the data. The physical upper limit for the
beat period ${\rm P}_{\rm B}$ is unknown.

When present around both radial modes, the two
triplets/quintuplets have nearly the same frequency spacings.
Specifically, the difference between the two spacings,
$|\Delta{\rm f}_2\! -\!\Delta{\rm f}_1|$ is always below $9\times
10^{-5}$\thinspace c/d. According to simulations of Alcock {\em
et~al.\/} (\cite{MACHO00}; their Fig.\thinspace 14), for MACHO
data such a frequency difference is statistically consistent with
no difference at all. In other words, triplets/quintuplets around
both radial modes have {\it the same frequency separation} within
accuracy of the data.

\subsection{Time Domain}

\begin{figure}[t]
\begin{center}
\includegraphics[width=10cm]{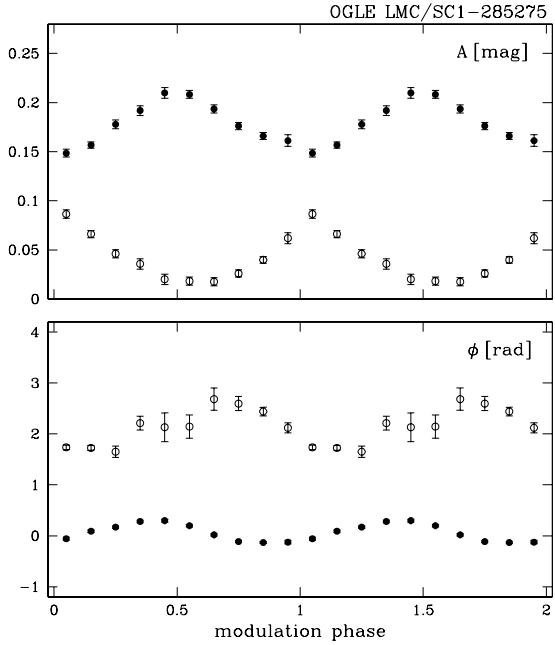}
\end{center}
\vskip -1.6truecm
\caption{Periodic modulation of FO/SO double mode Cepheid
LMC/SC1--285275. First and second overtones are displayed with
filled and open circles, respectively. Modulation period is
${\rm P}_{\rm B} = 891.6$\thinspace day.}
\label{fig3}
\end{figure}

The triplet/quintuplet frequency pattern can be interpreted as a
result of periodic modulation of amplitudes and/or phases of both
radial modes, with common period ${\rm P}_{\rm B} = 1/\Delta{\rm
f}$. In Fig.\thinspace\ref{fig3} we show this modulation for one
of the stars. The plot has been constructed by dividing the data
into 10 subsets, each spanning 10\% of the modulation cycle, and
then fitting radial modes and their linear combinations (Eq.(1))
to each subset separately. The resulting amplitudes and phases of
the modes are plotted {\it vs.} modulation phase.

Fig.\thinspace\ref{fig3} shows that both {\it amplitudes and
phases} undergo modulation. The amplitude variability is much
stronger for the second than for the first overtone (80\% {\it
vs.} 30\% change). {\it Minimum amplitude of one mode coincides
with maximum amplitude of the other mode}. Maximum phase of the
first overtone occurs just before maximum of its amplitude. For
the second overtone, maximum phase coincides with minimum
amplitude.

Characteristic pattern of periodic modulation displayed in
Fig.\thinspace\ref{fig3} is common to all Cepheids listed in
Table\thinspace\ref{table1}. The phenomenon is similar to the
Blazhko effect in RR~Lyrae stars ({\it e.g.} Kurtz {\em et~al.\/}
\cite{KUR00}). By analogy, variables of
Table\thinspace\ref{table1} can be called "Blazhko double mode
Cepheids".

\subsection{Stars with Unresolved Power Spectrum}

In 17 FO/SO double mode Cepheids (16\% of the sample) we find
after pre\-white\-ning a significant residual power unresolved
from the primary pulsation frequencies. This is a signature of
slow amplitude and/or phase variability, not resolved within the
length of available data ({\it i.e.} $\sim 2700$\thinspace days).

\begin{figure}[t]
\begin{center}
\includegraphics[width=10cm]{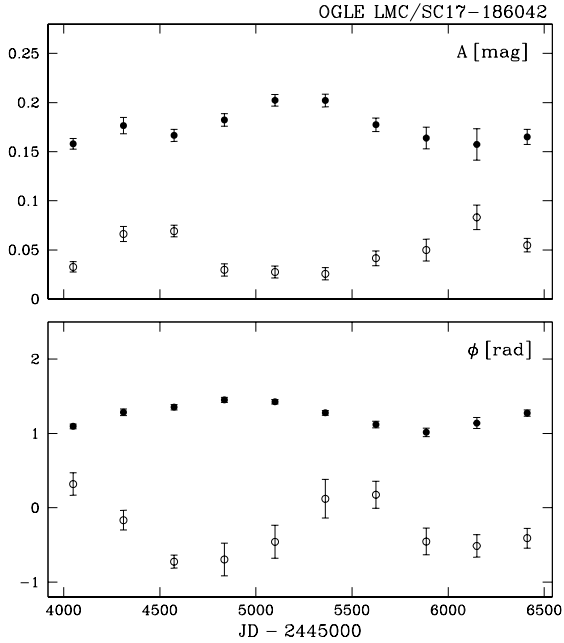}
\end{center}
\vskip -1.6truecm
\caption{Slow amplitude and phase variability of FO/SO double mode
Cepheid LMC/SC17--186042. First and second overtones are displayed
with filled and open circles, respectively.}
\label{fig4}
\end{figure}

For closer examination of this slow variability we have performed
for each star a similar analysis to that presented in
Fig.\thinspace\ref{fig3}. Specifically, we have divided the data
into 10 subsets, each spanning 10\% of the total timebase, and
then have fitted Eq.(1) separately to each subset. In some stars
we have detected only changes of pulsation phases, but in 10
variables we have also detected clear changes of the amplitudes.
In Fig.\thinspace\ref{fig4} we show typical example of such
behaviour. The pattern of amplitude and phase variability is very
much the same as that displayed in Fig.\thinspace\ref{fig3}. In
particular, the amplitudes of the two radial modes vary in
opposite directions. This similarity suggests, that at least some
of the unresolved FO/SO double mode Cepheids experience the same
type of periodic modulation as Cepheids listed in
Table\thinspace\ref{table1}, but on timescales longer than the
current data.

\begin{table}
\caption{Unresolved FO/SO Double-Mode Cepheids in the LMC}
\begin{center}
\small
\begin{tabular}{lcccclcc}
\hline
\noalign{\smallskip}
MACHO Star   & $P_1$\thinspace [day]
                      & $P_2$\thinspace [day]
                               &&& OGLE Star    & $P_1$\thinspace [day]
                                                         & $P_2$\thinspace [day]
                                                                  \\
\noalign{\smallskip}
\hline
\noalign{\smallskip}
 6.6934.67   & 0.9202 & 0.7400 &&& SC4--176400  & 1.1089 & 0.8951 \\
 11.9348.78  & 0.7378 & 0.5944 &&& SC4--220148  & 0.7408 & 0.5951 \\
 14.9098.35  & 0.7278 & 0.5865 &&& SC7--120511  & 1.2511 & 1.0037 \\
 15.10428.60 & 0.6518 & 0.5253 &&& SC10--204083 & 0.5263 & 0.4229 \\
 23.3061.82  & 0.5957 & 0.4810 &&& SC16--266808 & 1.3529 & 1.0810 \\
 24.2853.69  & 0.7088 & 0.5717 &&& SC17--186042 & 0.6110 & 0.4912 \\
 24.2855.80  & 0.5938 & 0.4783 &&& SC20--188572 & 1.0152 & 0.8145 \\
 47.2127.102 & 0.5785 & 0.4666 &&& SC21--12012  & 1.3415 & 1.0749 \\
 80.7079.62  & 1.3479 & 1.0764 &&&              &        &        \\
\noalign{\smallskip}
\hline
\end{tabular}
\end{center}
\label{table2}
\end{table}

\section{What Causes the Modulation ?}

Periodic amplitude and phase modulation in FO/SO double mode
Cepheids is by no mean a rare phenomenon. According to our
analysis its incidence rate is at least 19\% and perhaps as high
as 35\%.

Any model of modulated double mode Cepheids has to explain two
most basic properties of these stars:

\begin{itemize}
\item both radial modes are modulated with the same period.
\item amplitudes of the two modes change in opposite phase: maximum of
      one amplitude coincides with minimum of the other.
\end{itemize}

\noindent Two models have been proposed to explain Blazhko
modulation in RR~Lyrae stars: the oblique magnetic pulsator model
(Shibahashi \cite{SHI95}, \cite{SHI00}) and 1:1 resonance model
(Nowakowski \& Dziembowski \cite{NW01}). We will argue, that both
these models fail to account for modulation observed in the FO/SO
double mode Cepheids.

The oblique magnetic pulsator model assumes presence of a dipole
magnetic field, which is inclined to the rotation axis of the
star. The field introduces distortion to the radial pulsation. As
the star rotates, this distortion is viewed from different angles,
causing variation of the pulsation amplitude. However, in this
scenario both modes should display maximum amplitude
simultaneously, because they are both distorted in the same way.
This is not what we observe.

The model proposed by Nowakowski \& Dziembowski (\cite{NW01})
assumes resonant coupling of a radial mode to {\it nonradial}
modes of $\ell=1$. Such a mechanism ge\-ne\-ra\-tes in a natural
way a triplet of equally spaced frequencies. The beating of these
frequencies causes apparent amplitude and phase modulation.
Generalizing this model to coupling with modes of $\ell=2$, we can
also explain frequency quintuplets. However, in this resonance
model, modulation of each radial mode is an independent effect.
Therefore, there is no reason why both modes should be modulated
with the same period. There is also no reason why amplitude
variations of the modes should be in any particular phase relation
to each other (as is observed). Finally, the 1:1 resonance model
does not explain why in vast majority of cases modulation is
observed either in both radial modes or in none of them.

At this stage, the mechanism causing amplitude and phase
modulation of FO/SO double mode Cepheids remains unknown.
Nevertheless, the available observations already provide some
important clues. The fact that the two modes are always modulated
with the same period proves that their behaviour is not
independent. Both modes must be part of the same dynamical system.
The fact that high amplitude of one mode always coincides with low
amplitude of the other mode strongly suggests that energy transfer
between the two modes is involved. Thus, available evidence
clearly points towards some form of mode coupling in which both
radial modes take part.

\medskip

{\bf Acknowledgements}. This work has been supported by the Polish MNiSW Grant No. 1 P03D 011 30.



\end{document}